\documentclass[aps,english,prb,twocolumn,showpacs,superscriptaddress]{revtex4}
\usepackage{graphicx}
\makeatletter
\def\p@figure{Figure~}
\makeatother

\begin{document}

\title{Highly surface-active Ca(OH)$_2$ monolayer as a CO$_2$ capture material}

\author{V. Ongun \"{O}z\c{c}elik}\email{ongun@princeton.edu}
\affiliation{Andlinger Center for Energy and the Environment, Princeton University, New Jersey 08544 USA}
\affiliation{Department of Civil and Environmental Engineering, Princeton University, New Jersey 08544 USA}
\author{Kai Gong}
\affiliation{Andlinger Center for Energy and the Environment, Princeton University, New Jersey 08544 USA}
\affiliation{Department of Civil and Environmental Engineering, Princeton University, New Jersey 08544 USA}
\author{Claire E. White}\email{whitece@princeton.edu}
\affiliation{Andlinger Center for Energy and the Environment, Princeton University, New Jersey 08544 USA}
\affiliation{Department of Civil and Environmental Engineering, Princeton University, New Jersey 08544 USA}

\begin{abstract}

Greenhouse gas emissions originating from fossil fuel combustion contribute significantly to global warming, and therefore the design of novel materials that efficiently capture CO$_2$ can play a crucial role in solving this challenge.  Here, we show that reducing the dimensionality of bulk crystalline portlandite results in a stable monolayer material, named portlandene, that is highly effective at capturing CO$_2$. Based on theoretical analysis comprised of ab-initio quantum mechanical calculations and force-field molecular dynamics simulations, we show that this single-layer phase is robust and maintains its stability even at high temperatures. The chemical activity of portlandene is seen to further increase upon defect engineering of its surface using vacancy sites. Defect-containing portlandene is capable of separating CO and CO$_2$ from a syngas (CO/CO$_2$/H$_2$) stream, yet is inert to water vapor. This selective behavior and the associated mechanisms have been elucidated by examining the electronic structure, local charge distribution and bonding orbitals of portlandene. Additionally, unlike conventional CO$_2$ capturing technologies, the regeneration process of portlandene does not require high temperature heat treatment since it can release the captured CO$_2$ by application of a mild external electric field, making portlandene an ideal CO$_2$ capturing material for both pre- and post-combustion processes.

\end{abstract}

\maketitle

Although alternative energy technologies have advanced rapidly in recent decades, fossil fuel-based energy sources will play dominant roles in the majority of future energy scenarios.\cite{pacala2004stabilization} However, anthropogenic CO$_2$ and CO emissions originating from fossil fuel combustion are major contributors to the greenhouse gas effect that is altering the Earth's climate.\cite{Montzka2011} While CO$_2$ has a direct heat-trapping effect, CO indirectly contributes to global warming by reacting with hydroxyl radicals, subsequently leading to an increase in the concentration of O$_3$, methane and other trace gases in the atmosphere.\cite{Montzka2011, van2012greenhouse,Foster2017} There are conventional strategies targeting this problem by designing efficient combustion technologies that produce the same power output with lower carbon emissions, however the bottleneck for reducing CO$_2$ emissions is the separation/capture of CO$_2$ from the flue gas prior to storage. One major challenge of the separation/capture process is finding high performance adsorbents that can selectively capture CO$_2$ from the flue gas and release it easily during the regeneration process.\cite{lin2012silico} Therefore, the design of new capturing materials will improve the effectiveness of present CO$_2$ capture and storage technologies.

Currently, the main CO$_2$ capture technology used in practice relies on amine solutions that have limitations such as low capturing efficiencies, corrosion and toxicity.\cite{ferey2011hybrid, dutcher2015amine} Other well-known CO$_2$ capture approaches involve using solid materials such as nanotubes and metal-organic frameworks.\cite{cinke2003co, wang2008colossal, furukawa2010ultrahigh, ferey2011hybrid} For most of these cases, in order to recycle the capturing agent, the release of CO$_2$ requires the use of heat during the regeneration process. However, the ideal adsorbent material should work at low temperatures ($<$500K) to maximize the capture efficiency which is defined by the Department of Energy as capturing at least 90\% of CO$_2$ with an increase in the cost of electricity of no more than 10\%.\cite{doe, duan2010co} Previous studies have shown that bulk metal oxides and alkali-earth metal hydroxides (e.g., Ca(OH)$_2$ and Mg(OH)$_2$) are effective in adsorbing CO$_2$ at low to intermediate temperatures (between 300K and 1000K). Furthermore, it was shown that as a result of CO$_2$ capture, these bulk crystals form very stable carbonate products which require additional heat treatment (between 600K and 1400K) to release the captured CO$_2$.\cite{duan2010co} In terms of adsorbent efficiency, the Ca(OH)$_2$ and Mg(OH)$_2$ powders were seen to be very effective, where the amount of CO$_2$ adsorbed was inversely proportional to the domain size of the specimen \cite{lin2008magnesium, jahangiri2017effects} suggesting that the surface/volume ratio plays a critical role in the capturing process. Concomitantly, is was shown that certain 2D structures can be engineered to selectively adsorb gas molecules.\cite{lan2010doping, wang2011titanium, sun2013charge, oh2015selective, li2017understanding, tan2017borophene, sun2017electric, owuor2017lightweight} For instance, it was shown that titanium decorated graphene oxide can actively capture CO molecules \cite{wang2011titanium}, boron based structures borophene and  boron nitride (BN) \cite{tan2017borophene, sun2013charge}  can be utilized as CO$_2$ capturing media under external electric fields and 2D MoS$_2$\cite{sun2017electric} captures CO$_2$  when extra electrons are injected into the monolayer.

\begin{figure}
\includegraphics[width=7cm]{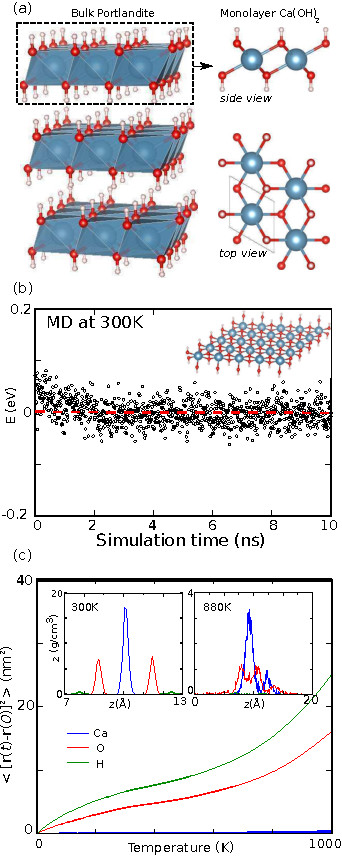}
\caption{Structure and stability of pristine portlandene (a) Structure of the monolayer portlandene exfoliated from the bulk portlandite crystal. Ca-O and O-H bond lengths are 2.38\AA~ and 0.97\AA, respectively. The lattice constant of the optimized hexagonal unit-cell is 3.62\AA.  In the ball-and-stick model Ca, O and H atoms are shown by blue, red and pink spheres, respectively.  (b) Variation of the total energy during the force-field MD simulation of portlandene at 300K. The energy values are normalized to primitive unit cell and the average value of the total energy is set to 0. (c) Mean square displacement of Ca, O and H atoms of portlandene over 10ns as temperature increases from 0K to 1000K. The insets show the mass density profiles at 300K and 880K.}
\label{fig1}
\end{figure}

\begin{figure*}
\includegraphics[width=12cm]{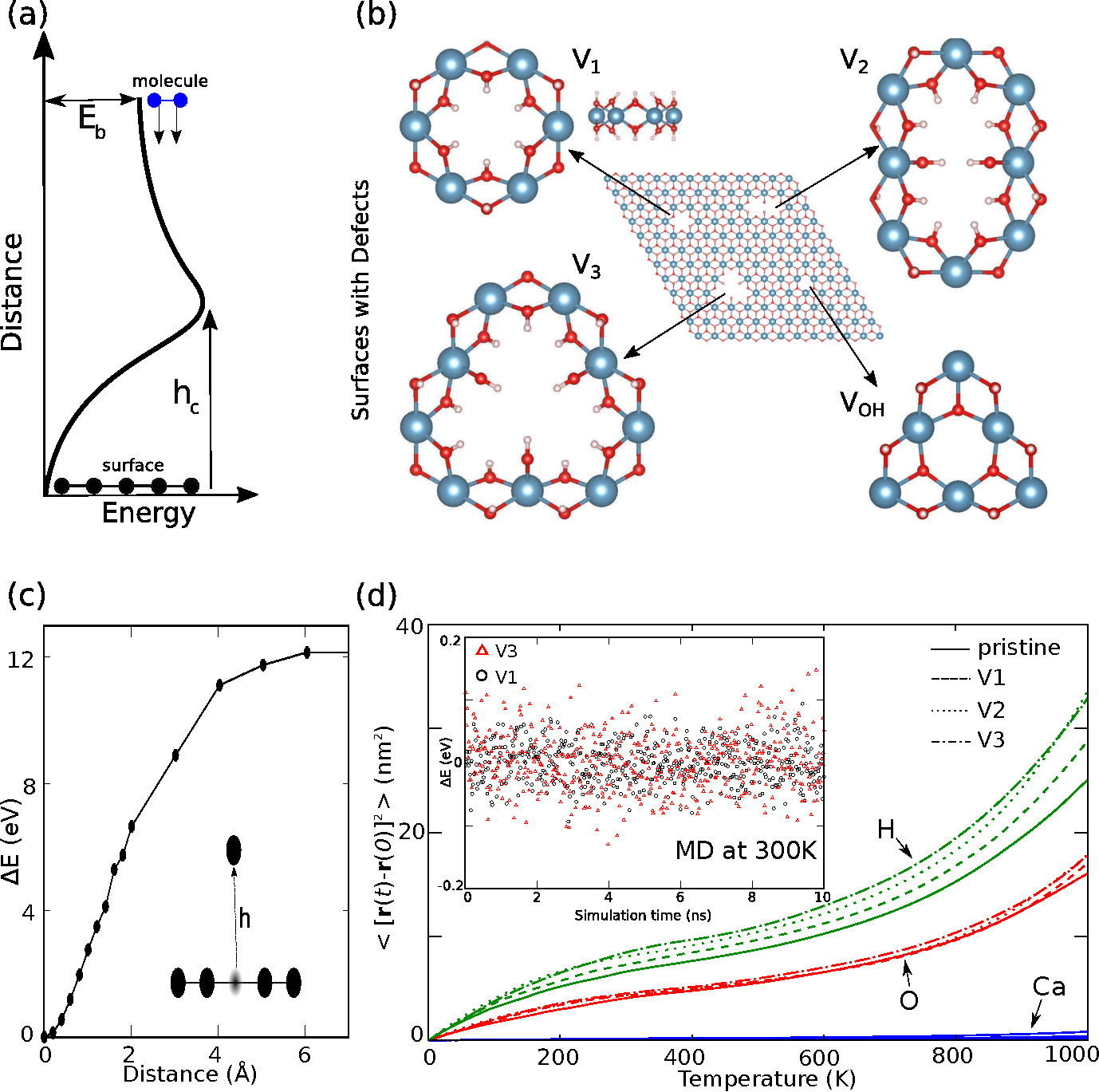}
\caption{(a) Schematic illustration showing the variation of total energy as a molecule approaches the portlandene surface. (b) Optimized geometries of defected structures.  (c) Variation of total energy as a Ca atom is pulled away from pristine portlandene. (d) Mean square displacement curves of pristine and defected portlandene calculated over 10ns as temperature increases from 0K to 1000K. The inset shows the variation of the total energy per primitive cell during the force-field MD simulation of portlandene with single (black) and triple (red) vacancies at 300K where mean value of energy is set to zero.}
\label{fig2}
\end{figure*}

In this letter, using density functional theory (DFT)  calculations and force-field molecular dynamics (MD) simulations, we theoretically predict that monolayer Ca(OH)$_2$ crystal (named as portlandene hereafter, in analogy to multilayer portlandite) is a thermodynamically stable material and can be used as an efficient CO$_2$ adsorbent over a wide range of temperatures (0-700K).  We show that the reactivity of portlandene with CO$_2$ can be increased by defect engineering its surface with vacancies of various sizes and these defects do not break the integrity of portlandene even at elevated temperatures. We reveal the fundamental mechanism behind the high reactivity of pristine/defected portlandene via monitoring the local charge distributions, bonding orbitals and the electronic structure of the material. Additionally, we show that monolayer portlandene can be used to separate CO/CO$_2$ mixture from the H$_2$ in syngas (CO$_2$/CO/H$_2$ gas stream), which is crucial for the precombustion process. Finally, we note that portlandene is inert to H$_2$O vapor, and therefore defect formation in portlandene and the subsequent CO$_2$/CO capture mechanism are not affected by the presence of water vapor.

\paragraph*{Structure and Stability of Portlandene}
Few-layered Ca(OH)$_2$ crystals have already been successfully synthesized at ambient conditions.\cite{aierken2015portlandite} Here, we start our analyses by showing that monolayer portlandene corresponds to a local energy minimum on the Born-Oppenheimer surface using various stringent stability tests. The optimized geometrical parameters of the monolayer structure are presented in ~\ref{fig1}(a). For this stable phase, a hexagonal unit cell consisting of one Ca atom and two OH units establishes a regular network where Ca atoms are located in the same plane and  OH units alternate on either side of the Ca plane. When perturbed from different directions, the structure returns back to its initial pristine geometry after relaxation using conjugate gradient methods. This is a result of the high formation energy of monolayer portlandene which we calculate as $E_f$=12.47eV using $E_f = E[Ca]+2E[OH]-E[Ca(OH)_2]$ where $E[Ca(OH)_2$], $E[Ca]$, $E[OH]$ are the ground state energies of the portlandene unit cell, single Ca atom and single OH group, respectively. Note that portlandene also has a cohesive energy of 4.44 eV per atom which is calculated using $E_c = (E[Ca]+2E[O]+2E[H]-E[Ca(OH)_2]) / 5 $. Although portlandene's high formation and cohesive energies indicate stability at 0K, we further examine its stability at ambient conditions using both ab-initio and force-field MD simulations. The final stable structure obtained after an ab-initio MD simulation of 5ps at room temperature is shown in the inset of ~\ref{fig1}(b). It is seen that portlandene remains stable even though the phonon modes soften at such temperatures. The MD simulations were further extended to 10ns using classical force-field methods. As shown by the variation of total energy in ~\ref{fig1}(b), the change in the energy of the system oscillates around an average value with an amplitude of 0.07 eV per unit cell. This is much less than the cohesive energy of the structure (4.44eV), indicating that these oscillations are too small to cause the monolayer to break apart.

Mean square displacement (MSD), which is a measure of the average distance an atom travels, provides further information about the comparative stability of materials at elevated temperatures. Here, we calculate MSD of portlandene as a function of temperature based on the MD trajectories using $MSD = <[\textbf{r}(t)-\textbf{r}(0)]^2>$ where $\textbf{r}(t)$ and $\textbf{r}(0)$ denote the position of an atom at time $t$ and $t=0$, respectively. ~\ref{fig1}(c) shows the MSDs of Ca, O, and H atoms in portlandene over 10ns as the temperature of the system was increased linearly from 0K to 1000K. It is observed that the slopes of the MSD curves increase suddenly above $\sim$720K. Since the slope of an MSD curve is closely related to the self-diffusion constant,\cite{nose1984molecular} these sudden increases indicate that O and H atoms diffuse away from their average positions rapidly above $\sim$720K. Such sudden changes in the slope of the MSD curve have also been observed during the crystallization process of a carbon nanotube-alkane nanocomposite,\cite{yang2011crystallization} glass transition in glassy and liquid polybutadiene,\cite{buchenau2014probing} and melting of graphene.\cite{kowaki2007radius} The inset of ~\ref{fig1}(c) shows the mass density profile (MDP) of each element in the structure over a 5ns MD simulation at 300K and 880K, respectively. The MDPs at 300K are clearly narrower than those at 880K. It is evident from the MDP plots that the structure is stable at room temperature, however it loses its integrity at 880K due to thermal effects. Our detailed MDP analysis shows that the OH groups start to dissociate from the main Ca-O sheet at a temperature of 780K, which aligns with the sudden change of the slope in the MSD curves and the experimentally measured decomposition temperature ($\sim$740K) of bulk Ca(OH)$_2$.\cite{zelic2002kinetic}

\paragraph*{Surface reactivity and defects}
Due to the ionic nature of the Ca-OH bonds in portlandene, the electrons shift toward the OH edges leading to an increase in the chemical activity of those sites.  When external CO$_2$ and CO molecules are introduced to the system, initially they are weakly attracted to the OH bonds via van der Waals (vdW) forces. This may cause the external molecules to be inert or become weakly trapped in a shallow energy minimum. However, if the external molecules get closer to the surface, they escape this shallow energy minimum and stronger interactions occur with the surface. To investigate the height at which these strong interactions become important, we bring the molecule closer to the surface in small incremental steps as illustrated in ~\ref{fig2}(a) and perform a structure optimization for each step until we come to a distance where a strong attractive interaction sets in between the molecule and the surface.  We define this distance as the critical adsorption distance (h$_c$) and as soon as the molecule gets closer than h$_c$ to portlandene, it is adsorbed to the surface at the OH sites. Upon adsorption, CO$_2$ and CO molecules neither dissociate into their constituent atoms nor destroy the portlandene structure. Instead, they are trapped on the portlandene surface as molecules. The binding energy and critical adsorption distance of CO$_2$ on portlandene are 120meV and 2.0\AA, respectively. CO always weakly interacts with portlandene and has a binding energy of 7meV. Similarly, we also note that the H$_2$ molecule is totally inert to the surface. The binding energies are calculated using $E_b =   E[portlandene] + E[molecule] - E[molecule+portlandene]$. The large difference between the binding energies of CO$_2$ and CO/H$_2$ molecules suggests that portlandene can be used as a material which can separate CO$_2$ from CO/H$_2$, since CO$_2$ atoms will be captured by the surface in a mixed CO/CO$_2$/H$_2$ gas flow. 

The reactivity of portlandene can be further increased by decorating its surface with Ca vacancy defects which are very common and easy to create in monolayer materials during growth and exfoliation processes.\cite{banhart2010structural, bjorkman2013defects}  Defect formation in pristine structures might seem undesirable at first sight, however controlled defect engineering results in novel materials via modification of the electronic and chemical properties. It has been shown that when a vacant site forms, the reactivity of the atoms surrounding the vacancy will increase due to their reduced coordination state.\cite{boukhvalov2009chemical, carlsson2009two} As a result of this increased chemical activity, these sites can be used for dissociation or adsorption of foreign molecules depending on the material and molecule of interest.\cite{zhu2007sublimation, kostov2005dissociation,gurel2014dissociative} Here, we create vacancy defects in portlandene by the removal of Ca atoms, resulting in the optimized geometries shown in ~\ref{fig2}(b). Note that the triple coordination of OH unit (to Ca) drops to double in the single vacancy defect (V1), whereas double (V2) and triple (V3) vacancies lead to both singly and doubly coordinated OH units. An OH vacancy does not change the coordination number of other OH units. 

\begin{figure}
\includegraphics[width=8cm]{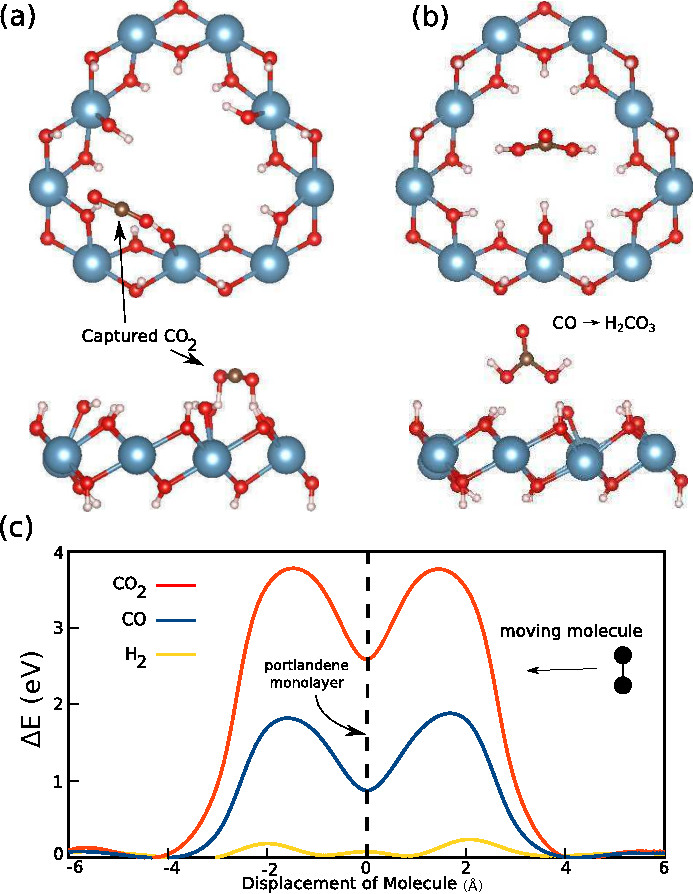}
\caption{(a) Top and side views of the optimized geometry of a CO$_2$ molecule adsorbed on defected portlandene. (b) Same for CO. Note that instead of adsorption, CO pulls two OH units from the surface to form H$_2$CO$_3$. (c) Penetration barriers of CO$_2$, CO and H$_2$ molecules through portlandene with V3 vacancy defect. }
\label{fig3}
\end{figure}

Before investigating the effect of vacancy sites and the reduced coordination numbers on the chemical reactivity of portlandene, we initially evaluate their stabilities. Defect formation energy is the prime criterion that determines the stability of a defect at its ground state. It is defined as the energy required to completely remove an atom from the monolayer or let it randomly migrate to another site on the flake.\cite{gillan1989calculation, kittel} Variation of total energy during controlled defect formation in portlandene can be modeled by pulling an atom out of the monolayer and recalculating the ground state energy at each intermediate step as the atom leaves the surface. ~\ref{fig2}(c) shows that as a Ca atom moves further away from the surface the total energy gradually increases and saturates at a constant value of 12.10eV once the atom is completely detached. This large energy difference between the initial and final configurations gives the pulling energy ($E_p$) required to remove a single Ca atom from the pristine structure and is defined as $E_p=E[V1]+E[Ca]-E[portlandene]$ where $E[V1]$ is the total energy of portlandene with a single Ca vacancy.\cite{ozccelik2013self} The defect formation energy can be computed by subtracting the cohesive energy of portlandene from E$_p$  which gives 7.66 eV for a single Ca vacancy defect. Similarly, the formation energies of double and triple Ca vacancies are 7.33 eV and 7.19 eV, respectively, indicating that the vacancy defects are stable and do not cause portlandene to lose its integrity spontaneously at the ground state. 

To evaluate the effect of elevated temperature on vacancy defects we compare the MSD curves of pristine and defected structures where each structure is heated from 0K to 1000K in 5ns as shown in  ~\ref{fig2}(d). The MSD curves for pristine, V1, V2 and V3 structures show similar behavior for Ca and O atoms whereas defects induce a slight shift in the MSD of the H atoms. Nevertheless, it is clear that the introduction of vacancies only slightly modifies the MSD of portlandene which suggests similar stability as compared to the pristine structure at elevated temperatures. The inset of ~\ref{fig2}(d) shows the variation of the total energy for V1 and V3 structures during the MD simulations performed at 300K for 10ns. During these simulations the total energy oscillates around an average value with amplitudes of 0.06eV, 0.10eV, 0.14eV for V1, V2, and V3 structures, respectively. These oscillations are much smaller than the defect formation energies calculated earlier and  therefore indicate that the integrity of the structure is not affected by the presence of vacancies. Finally, we note that although the monolayer does not dissociate into its constituent atoms, vacancies do cause structural reconstruction around the defect site such that the OH units adjacent to the vacancies bend toward each other, reducing the overall size of the vacancy.

Having analyzed the stability of Ca vacancy defects, we return to our initial quest regarding the reactivity of defect-containing portlandene with CO$_2$ and CO molecules. Following a similar procedure to what we carried out for the pristine structure, we calculate the binding energies and critical adsorption heights for each molecule on the defected portlandene structures. For the case of the CO$_2$ molecule, when the CO$_2$ - portlandene distance drops below 2.3\AA, an attractive interaction sets in. CO$_2$ is then drawn toward the dangling OH units around the defect until it physisorbs to the surface. OH units which are in two-fold (V1 defect) or one-fold (V2 and V3 defects) coordination are strongly bonded to the Ca atoms, however they are seen to tilt toward the CO$_2$ molecule as a result of the attractive interaction. In the optimized final configuration, the CO$_2$ molecule is trapped between two adjacent OH units around the vacancy defect as shown in ~\ref{fig3}(a).  This binding mechanism takes place for the V1, V2 and V3 structures with binding energies of 0.25eV, 0.54eV and 0.55eV, respectively. Here we note that the difference in binding energies can be attributed to the fact that for V1, the chemically active OH units are doubly coordinated (bonded to two Ca atoms) whereas for V2 and V3 they are singly coordinated (bonded to one Ca atom). Therefore, OH units around V2 and V3 defects have more free electrons to interact with CO$_2$ as compared to the OH units around the V1 defect. These calculated binding energies are significantly larger than the adsorption energies of CO$_2$ on other 2D materials which were reported as 0.01eV for MoS$_2$, 0.18eV for borophene, 0.13eV for BN sheet and 0.06eV for BN nanotube. \cite{tan2017borophene, sun2013charge, sun2017electric}

Although the CO molecule only weakly bonds to the pristine portlandene surface, the surface attraction increases significantly in the presence of defects. The attraction begins as soon as the distance between CO and the OH units associated with the defects decreases below 1.8\AA.~ The CO molecule reorients itself such that its C end points toward an OH unit which eventually detaches from the surface. This is followed by the detachment of another adjacent OH unit and the formation of carbonic acid, H$_2$CO$_3$. H$_2$CO$_3$ then freely migrates away from the defect region as shown in ~\ref{fig3}(b). For a single vacancy, the final defected portlandene + H$_2$CO$_3$ configuration is more favorable than the initial defected portlandene + CO system by 6.1eV per unit cell. This energy difference increases to 6.4eV for portlandene with double and triple vacancies. Here we note that traditionally, CO capture and filtration is achieved by catalytic oxidation of CO into CO$_2$, which requires an additional step to prevent the emission of CO$_2$ into the atmosphere (possibly by CO$_2$ storage). However, our results show that defected portlandene can capture CO by  releasing H$_2$CO$_3$ which can easily be collected and removed from the stream. It should be noted that H$_2$CO$_3$ can also dissociate into CO$_2$ and H$_2$O, and in this case the released CO$_2$ will be subsequently collected by the defect sites in portlandene.

In order for CO$_2$ and CO to interact with defects, the distance between the molecules and OH units must be lower than a critical value, as discussed above. Thus, it is crucial to investigate if the molecules can penetrate the defect sites through the middle of the vacancy without feeling the attractive forces from the OH units around the vacancy edges. To investigate this penetration mechanism, we calculate the energy barriers associated with the external molecules passing through the vacancies shown in ~\ref{fig2}(b). Starting from a large distance, we bring the molecule sequentially closer to the middle of the defect and calculate the total energy of the system for each molecule - monolayer distance. During this process, the coordinates of all atoms are relaxed except for the Ca atoms at the corners of the portlandene supercell to prevent the surface from sliding. As shown in ~\ref{fig3}(c) for V3, both CO$_2$ and CO molecules have high penetration barriers of 3.6eV and 1.9eV, respectively, whereas H$_2$  can easily penetrate through the vacancy due to its compact size. These high energy barriers show that CO$_2$/CO molecules cannot penetrate through the defect and instead will interact with the defect site provided that their distance from the monolayer is below the critical value. CO$_2$ and CO are seen to remain on the same side of the defect until they diffuse closer to the chemically active OH units, at which point they will interact with the OH units as described above.

\begin{figure*}
\includegraphics[width=14cm]{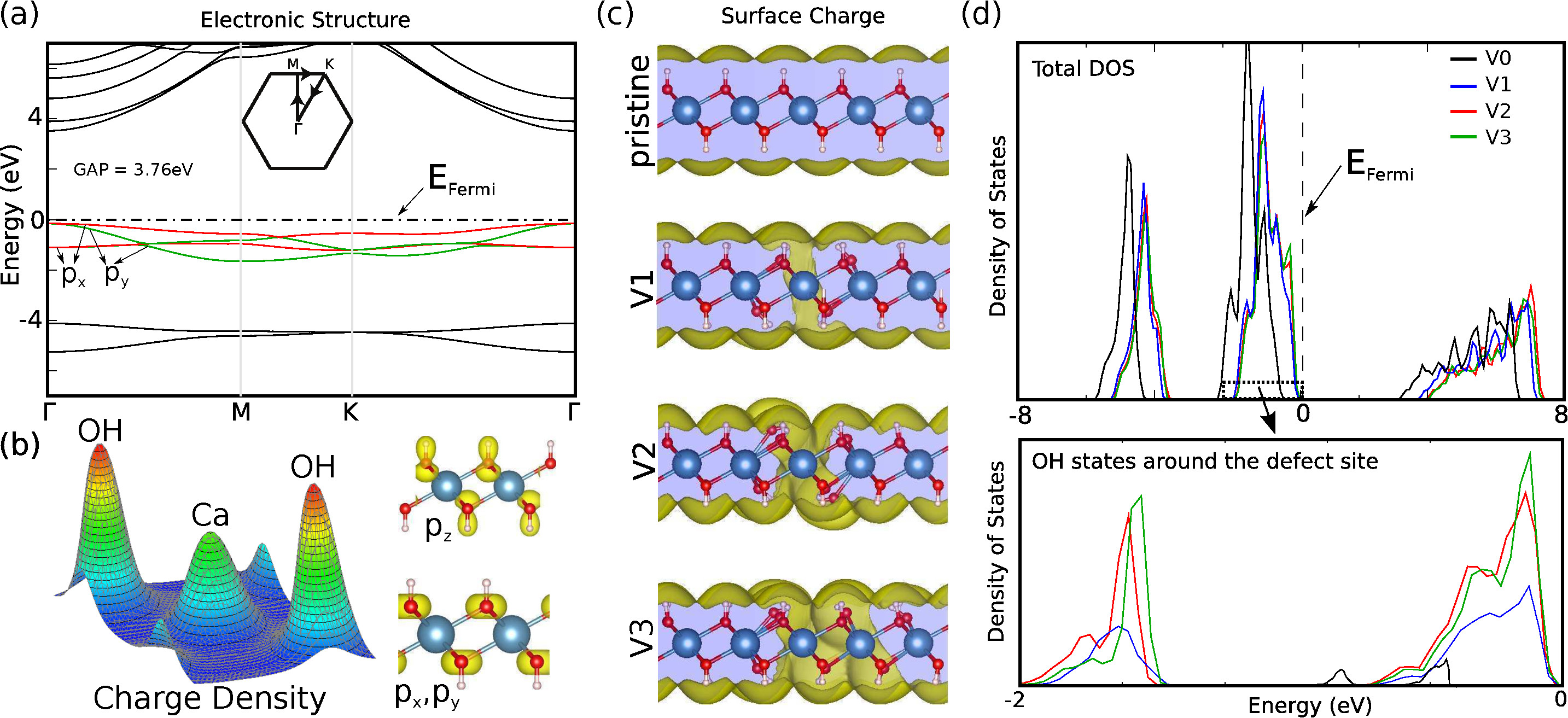}
\caption{(a) Electronic band diagram of pristine portlandene. (b) Left: Distribution of electron density on the Ca-OH plane. Right: (px,py) and pz orbitals calculated at VBM and CBM, respectively. (c)  Total charge densities of pristine and defected portlandene. (d) Top: Total density of states (DOS) of pristine and defected portlandene. Bottom: Local DOS of OH units at the valence band edge.}
\label{fig4}
\end{figure*}

\paragraph*{Electronic Structure}
The increased chemical reactivity of defected portlandene manifests itself in the electronic structure of this monolayer. Pristine portlandene is a non-magnetic material with a direct band gap of 3.76eV as shown in ~\ref{fig4}(a). Its valence band maximum (VBM) and conduction band minimum (CBM) are both at the zone center ($\Gamma$ point). The pristine structure has zero net magnetic moment and a negligible spin-orbit coupling effect, except for 0.02meV splitting at the VBM. Distribution of electronic charge on the monolayer is shown in ~\ref{fig4}(b). Accordingly, the net charge transfer from a Ca atom to an adjacent OH unit is 0.89e and the electrons accumulate on the OH units.  The band decomposed charge densities presented in ~\ref{fig4}(b) show that the valence bands near the Fermi level are dominated by p$_x$ and p$_y$ orbitals of the O atoms while the CBM has p$_z$ contributions at the $\Gamma$ point. The existence of p$_z$ orbitals is a crucial indicator of high reactivity of the monolayer structures since the electrons in the p$_z$ orbital form a sheet of electron cloud with low ionization energies on both sides of the Ca(OH)$_2$ layer, as shown in ~\ref{fig4}(c). This electron cloud is prone to initiate chemical reactions with external molecules and the thickness of the electron cloud increases in the presence of vacancy defects. In addition to the electron layer forming on both sides of the monolayer, valence electrons also accumulate near the defected region. 

\begin{figure}
\includegraphics[width=7cm]{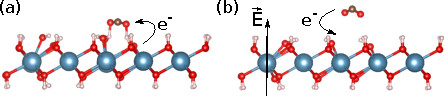}
\caption{Side view of the portlandene + CO$_2$ system at (a) zero electric field and (b) when an external electric field of 0.2V/\AA~ is applied.}
\label{fig5}
\end{figure}

To further analyze the effect of defects on reactivity, in ~\ref{fig4}(d) we present the variation of the density of states as defects are introduced. The electronic states at the VBM and CBM are dominated by OH states whereas higher conduction bands are dominated by Ca atoms. Upon creation of a single vacancy, the structure gains a net magnetic moment of 1.86 $\mu_B$ and the CBM shifts to a higher energy value of 3.8eV. The shift in the CBM continues as the second (V2) and third (V3) defects are created. The lower panel of ~\ref{fig4}(d) shows the contribution from the OH electrons at the defect site to the valence bands just below the Fermi level.   The contribution from the OH states to the VBM increases after the creation of the first and second vacancies, which is attributed to the reduced coordination of the OH units. Since these valence electrons can move into an excited state by an external stimulus (by an approaching CO$_2$ molecule in our case), further charge accumulation on the dangling OH units indicates higher reactivity. The V2 and V3 structures have similar characteristics for the VBM due to the fact that they both have singly and doubly coordinated OH units. Nevertheless, contributions from OH states to the VBM are higher for V3 compared with V2. As a final remark we note that defected portlandene can release CO$_2$ under an external electric field. This is crucial since an ideal capturing material should also easily release CO$_2$ for regeneration when needed. It is well-known that an external field can alter the chemical reactivity of materials by modulating their charge distributions.\cite{sun2013charge,tan2017borophene, sun2017electric} Here, we observe that when an electric field of 0.2V/\AA~ is applied perpendicular to the monolayer, the charge transfer from the monolayer to the CO$_2$ is reversed and CO$_2$ desorbs from the surface as shown in ~\ref{fig5}. This initial result suggests that the binding energy and adsorption location of CO$_2$ on portlandene can be fine-tuned by adjusting the magnitude and direction of the external electric field, however further research is necessary to investigate the fundamentals of the release mechanism which is beyond the scope of this paper.

\paragraph*{Conclusion}
In conclusion, we have shown that monolayer portlandene can be used as a CO$_2$ capturing material even at low temperatures and its reactivity can be further improved by the presence of vacancy defects. Our ab-initio DFT and force-field MD calculations revealed that defects remain stable in the portlandene structure even at elevated temperatures up to 720K. Apart from capturing CO$_2$, defect-free portlandene can also act as a material which can successfully separate CO$_2$ from the CO/H$_2$ gas mixture. Furthermore, for portlandene containing defects, H$_2$ molecules can easily penetrate the vacancy sites whereas CO chemically interacts with the edges of these sites to form H$_2$CO$_3$. We showed that defects alter the electronic structure and charge distribution on the portlandene surface which is crucial for the increased reactivity of this monolayer material. We finally demonstrated that the adsorbed CO$_2$ molecules can easily detach from the surface by applying an external perpendicular electric field. Our theoretical predictions will lead to future experiments validating and exploring the capture properties of portlandene, and will pave the way for novel CO$_2$ capture technologies. 

\paragraph*{Methods}
Our predictions were obtained from spin polarized DFT calculations and force-field MD calculations at elevated temperatures. DFT calculations were performed within the generalized gradient approximation (GGA) including van der Waals corrections.\cite{grimme2006semi} We used projector-augmented wave potentials,\cite{blochl94} and the exchange-correlation potential was approximated with Perdew-Burke-Ernzerhof (PBE) functional.\cite{pbe} Vacancy defects were represented by using the supercell method, whereby  defected sites in a $(8 \times 8)$ supercell repeat themselves periodically. The size of this supercell was tested to be sufficiently large to avoid defect-defect coupling. The Brillouin zone was sampled by 9x9x1 \textbf{k}-points in the Monkhorst-Pack scheme where the convergence in energy as a function of the number of \textbf{k}-points was tested. A plane-wave basis set with energy cutoff value of 550 eV was used. Atomic positions were optimized using the conjugate gradient method, where the total energy and atomic forces were minimized. The energy convergence value between two consecutive steps was chosen as $10^{-5}$ eV. A maximum force of 0.01 eV/\AA~ was allowed on each atom. Ab-initio finite temperature MD calculations were performed where the time step was taken as 2.5 fs and the atomic velocities were renormalized to the temperature set at T = 300 K at every 40 time steps using an NVT ensemble. Ab-initio calculations were carried out using the VASP software.\cite{vasp} The stability of the intact and defective structures obtained from the DFT calculations were further evaluated using force-field MD simulations as implemented in the Atomistix ToolKit (ATK) package.\cite{toolkit2014quantumwise, griebelbook, griebel2004molecular} MSD curves and MDP plots were obtained by using ATK classical potential models\cite{dolado2007molecular,litton2001modeling, feuston1990onset, su2004role} which were specifically developed to cover material systems containing Ca, O, H and Si. An MD run of 10$^7$ steps was performed on each structure at a temperature of 300K using NVT ensemble and the Nose-Hoover thermostat.\cite{martyna1992nose} A time step of 1fs was adopted, which gives a total simulation time of 10 ns for each structure. 

\paragraph*{Acknowledgment:}

This work was supported by the Andlinger Center for Energy and the Environment (Princeton University). The participation of KG in this work was supported by Grant No. 1362039 from the National Science Foundation. The calculations were performed on computational resources supported by the Princeton Institute for Computational Science and Engineering (PICSciE) and the Office of Information Technology’s High Performance Computing Center and Visualization Laboratory at Princeton University.

\bibliography{arxiv.bbl}

\end{document}